\documentstyle[aps,preprint]{revtex}
\begin{document}
\preprint{gr-qc/9503015}
\draft
\title{Self-adjoint extensions and Signature Change}
\author{I.L.
Egusquiza\cite{emUPV}}
\address{Dpt. of Theoretical Physics\\
University of the Basque Country\\
Apdo. 644 P.K. - 48080 BILBAO\\
SPAIN}
\date{8 March 1995}
\maketitle
\begin{abstract}
We study the selfadjoint extensions of the spatial part of the D'Alembert
operator in a spacetime with two changes of signature. We identify a set of
boundary conditions, parametrised by $U(2)$ matrices, which correspond to
Dirichlet boundary conditions for the fields, and from which we argue against
the suggestion that regions of signature change can isolate singularities.
\end{abstract}  
\section{Introduction} In a recent preprint, Alty and Fewster \cite{alfe}
reexamine the problem of signature changing gravitational backgrounds in terms
of the existence of self-adjoint extensions of the Laplace-Beltrami operator.
Even though this work does not answer the question of whether quantum theory
makes sense in this kind of background, it has the merit of addressing the
issue
of well-posedness of the Klein-Gordon equation in them.

We shall extend their work to a spacetime with two discontinuous changes of
signature and identify  among the possible self-adjoint extensions a ($U(2)$)
subset for which a Casimir effect is apparent. In the light of this result the
Casimir effect itself will be reinterpreted.

 The conclusion follows that a kleinian region in an otherwise lorentzian
spacetime is unstable under the presence of a scalar quantum field, thus
providing an argument against the suggestion that regions of signature change
can isolate singularities.

 It must be pointed out that a fairly heated debate concerning junction
conditions for solutions of the Klein-Gordon equation in such signature
changing backgrounds is still under way \cite{debate,alty}. This short
note does not concern itself with the issue in the manner debated,
since the stress, following \cite{alfe}, is not on the existence of weak
solutions, but rather on the selfadjointness of the character-changing
differential operator in an $L^2({\bf R})$ context.

\section{The metric and the self adjoint extensions.}
Consider a metric of the form
$${\rm d} s^2= {\rm d}t^2 + (1-2\theta(x+a) + 2\theta(x-a)){\rm d}x^2\,.$$
The spatial ($x$) part of the associated Dalambertian is evidently well-defined
on the domain of infinitely continuously differentiable functions with compact
support not including the points $x=\pm a$. We look for self-adjoint extensions
of this Laplace-Beltrami operator $\Delta$ in the Hilbert space of square
integrable functions on the real line. As is well known \cite{Weid}, this will
be achieved if we examine the kernels of $\Delta^*\mp i$ and find them to be
unitarily equivalent. That is, we have to look for normalizable solutions
($\psi$) to the problem
$$\langle\psi|\Delta\varphi\rangle=\mp i\langle\psi|\varphi\rangle\,,$$
for all $\varphi\in C_0^\infty({\bf R}\backslash\{-a,a\})$, and the standard
measure in $\bf R$.

Orthonormal bases for each of these two subspaces are readily found
to be four-dimensional, thus proving the operator to be amenable to
self-adjoint
extension. The family of self-adjoint extensions is parametrized by elements of
the $U(4)$ group, as follows. Let the bases for ${\rm Ker}(\Delta^*\mp i)={\cal
K}^\pm$ be
\begin{eqnarray}
\chi_1^+(x) & = &2^{1/4} e^{a/\sqrt{2}} e^{\lambda
x}\left(1-\theta(x+a)\right)\,,\nonumber\\
\chi_2^+(x) & = &{{2^{1/4}\sinh{\lambda^3
x}\left(\theta(x+a)-\theta(x-a)\right)}\over{\sqrt{ \sinh(\sqrt{2}
a)-\sin(\sqrt{2}a)}}} \,,\nonumber\\ \chi_3^+(x) & = &{{2^{1/4}\cosh{\lambda^3
x}\left(\theta(x+a)-\theta(x-a)\right)}\over{\sqrt{ \sinh(\sqrt{2}
a)+\sin(\sqrt{2}a)}}} \,,\nonumber\\ \chi_4^+(x) & = &2^{1/4} e^{a/\sqrt{2}}
e^{-\lambda x}\theta(x-a)\,,\nonumber\\
\end{eqnarray}
and
\begin{eqnarray}
\chi_1^-(x) & = &2^{1/4} e^{a/\sqrt{2}} e^{-\lambda^3
x}\left(1-\theta(x+a)\right)\,,\nonumber\\
\chi_2^-(x) & = &{{2^{1/4}\sinh{\lambda
x}\left(\theta(x+a)-\theta(x-a)\right)}\over{\sqrt{ \sinh(\sqrt{2}
a)-\sin(\sqrt{2}a)}}} \,,\nonumber\\
\chi_3^-(x) & = &{{2^{1/4}\cosh{\lambda
x}\left(\theta(x+a)-\theta(x-a)\right)}\over{\sqrt{ \sinh(\sqrt{2}
a)+\sin(\sqrt{2}a)}}} \,,\nonumber\\
\chi_4^-(x) & = &2^{1/4} e^{a/\sqrt{2}}
e^{\lambda^3 x}\theta(x-a)\,,\nonumber\\
\end{eqnarray}
where $\lambda=\exp(i\pi/4)=(1+i)/\sqrt{2}$.

The functions in ${\cal K}^+$ can be thus written as
\begin{equation}
\psi_+=\left(\chi_1^+,\chi_2^+,\chi_3^+,\chi_4^+\right)
\pmatrix{\alpha\cr\beta\cr\gamma\cr\delta\cr}\,,
\end{equation}
for some parameters $\alpha$ to $\delta$.

For each self-adjoint extension, parametrised by a unitary transformation $U$,
the latter can be written as a unitary matrix acting on the vector of
parameters, thus having that the domain of the self-adjoint extension is given
by the addition (to the  infinitely continuously differentiable functions with
compact support not including the points $x=\pm a$) of the functions
$$\left(\chi_1^+,\chi_2^+,\chi_3^+,\chi_4^+\right)
\pmatrix{\alpha\cr\beta\cr\gamma\cr\delta\cr} +
\left(\chi_1^-,\chi_2^-,\chi_3^-,\chi_4^-\right) U
\pmatrix{\alpha\cr\beta\cr\gamma\cr\delta\cr}\,.$$

The explicit expressions of $\chi^\pm_i$ allow us to write $\psi(x)$, whence
$\psi(-a^+)$ and $\psi(a^-)$ can be readily computed. Were we to demand that
both these values equal zero for any set of parameters, we would see that this
is achievable provided we restrict ourselves to a $U(2)$ subgroup of the
original $U(4)$ possible self-adjointness conditions; namely, $U$ has to be
given by $$U=\pmatrix{a_{11}&0&0&a_{14}\cr0&e^{2i\nu}&0&0\cr0&0&-e^{2i\mu}&0\cr
a_{41}&0&0&a_{44}\cr}\,,$$
where
$$\pmatrix{a_{11}&a_{14}\cr a_{41}&a_{44}\cr}$$
is a $U(2)$ matrix, and $\nu$ and $\mu$ are given explicitly by
$$\tan(\nu)=-\coth(a/\sqrt{2})\,\tan(a/\sqrt{2})\,,$$
and
$$\tan(\mu)=-\tanh(a/\sqrt{2})\,\tan(a/\sqrt{2})\,.$$

It is immediate to observe (from the very block structure of the $U$ matrix,
for example) that the initial data for the derivatives of $\psi$ at $\pm a^\mp$
are disconnected from the data at $\pm a^\pm$: the region with the ``correct",
Minkowskian signature decouples from the Kleinian region, but mantaining at the
same time the necessary self-adjointness of the Laplace operator.

It is thus clear that it would be consistent, in attempting to quantize a
scalar field in the signature-changing background under consideration, to
impose Dirichlet boundary conditions on the signature-changing  walls,
thus leading to a Casimir effect that would correspond to an attraction between
the walls \cite{fulling,birrdavies}.

 In this context the Casimir effect can be then understood as being due to the
following complex of facts: whereas a positive frequency separation is feasible
(with Dirichlet boundary conditions) in the Minkowskian region, the operator
problem in the Kleinian region has no means to be understood as a propagation
problem. Therefore, positive and negative energies are completely mixed in the
Kleinian region, thus leading to a negative energy density with respect to the
Minkowskian part of spacetime. As a consequence, the collapsing motion of the
walls is energetically favourable.

\section{Conclusions}

We can now argue against the suggestion that whenever naked singularities
appear
in stringy gravity these will be isolated from real observers by a region of
signature change. For an (important) example of such behaviour, see
\cite{perryteo}. It
seems from the arguments above that imposing Dirichlet boundary conditions on
the
fields at the signature changing boundaries we should get a sensible quantum
field theory (insofar as that is at all possible, of course) in the signature
changing gravitational background under consideration. The presence of these
quantum fields would then make these signature changing boundaries unstable,
and
the region of Kleinian signature would either collapse, thus unstripping the
naked singularity of its hypothetical veil, or it would expand without bound,
reproducing the nasty effects of an unchecked nakedness.

No comment has been made on the classical/quantum stability
of cosmological solutions with signature change, but it can now be noticed that
i) classical stability has already been studied by other authors
\cite{haywardcosmo}; ii) in such contexts there would normally be just one
signature changing spacelike surface.

Another point has to be raised: in this note we have exclusively dealt with a
discontinuous signature change. Another long debate can also be found in the
literature concerning the respective merits of continuous versus discontinuous
signature change, but in our case this will not be so very relevant, since the
results put forward, namely, that the generalized Laplace operator admits a
self-adjoint extension which corresponds to Dirichlet boundary conditions on
the signature-changing wall, still hold, as can be checked with ease for the
${\rm d}t^2 + x {\rm d}x^2$ metric, for instance.

\end{document}